\documentclass[10pt,a4paper]{article}
\usepackage[utf8]{inputenc}
\usepackage[T1]{fontenc}
\usepackage{amsmath}
\usepackage{amsfonts}
\usepackage{amssymb}
\usepackage{graphicx}
\usepackage{fullpage}

\begin{document}

\title{ Synch: A framework for concurrent data-structures and benchmarks}
\author{Nikolaos D. Kallimanis\\
 Institute of Computer Science\\
 Foundation for Research and Technology-Hellas (FORTH)}
\date{~~~}
\maketitle
	
\section{Summary}

The recent advancements in multicore machines highlight the need to simplify concurrent programming in order to leverage their computational power. One way to achieve this is by designing efficient concurrent data structures (e.g. stacks, queues, hash-tables, etc.) and synchronization techniques (e.g. locks, combining techniques, etc.) that perform well in machines with large amounts of cores. In contrast to ordinary, sequential data-structures, the concurrent data-structures allow multiple threads to simultaneously access and/or modify them.

Synch~\cite{SynchFramework} is an open-source framework that not only provides some common high-performant concurrent data-structures, but it also provides researchers with the tools for designing and benchmarking high performant concurrent data-structures. The Synch framework contains a substantial set of concurrent data-structures such as queues, stacks, combining-objects, hash-tables, locks, etc. and it provides a user-friendly runtime for developing and benchmarking concurrent data-structures. Among other features, the provided runtime provides functionality for creating threads easily (both POSIX and user-level threads), tools for measuring performance, etc. Moreover, the provided concurrent data-structures and the runtime are highly optimized for contemporary NUMA multiprocessors such as AMD Epyc and Intel Xeon.

\section{Statement of need}

The Synch framework aims to provide researchers with the appropriate tools for implementing and evaluating state-of-the-art  concurrent objects and synchronization mechanisms. Moreover, the Synch framework provides a substantial set of concurrent data-structures giving researchers/developers the ability not only to implement their own concurrent data-structures, but to compare with some state-of-the-art data-structures. The Synch framework has been extensively used for implementing and evaluating concurrent data-structures and synchronization techniques in papers, such as \cite{AKD12,FK2011,FK2012,FK2014,FK2017,FKR2018}.

\section{Provided concurrent data-structures}

The current version of the Synch framework provides a large set of high-performant concurrent data-structures, such as combining-objects, concurrent queues and stacks, concurrent hash-tables and locks. The cornerstone of the Synch framework are the combining objects. A Combining object is a concurrent object/data-structure that is able to simulate any other concurrent object, e.g. stacks, queues, atomic counters, barriers, etc. The Synch framework provides the PSim wait-free combining object \cite{FK2011,FK2014}, the blocking combining objects CC-Synch, DSM-Synch and H-Synch \cite{FK2012}, and the blocking combining object based on the technique presented in \cite{Oyama99}. Moreover, the Synch framework provides the Osci blocking, combining technique \cite{FK2017} that achieves good performance using user-level threads.

In terms of concurrent queues, the Synch framework provides the SimQueue \cite{FK2011,FK2014} wait-free queue implementation that is based on the PSim combining object, the CC-Queue, DSM-Queue and H-Queue \cite{FK2012} blocking queue implementations based on the CC-Synch, DSM-Synch and H-Synch combining objects. A blocking queue implementation based on the CLH locks \cite{C93,MLH94} and the lock-free implementation presented in \cite{MS96} are also provided. In terms of concurrent stacks, the Synch framework provides the SimStack \cite{FK2011,FK2014} wait-free stack implementation that is based on the PSim combining object, the CC-Stack, DSM-Stack and H-Stack \cite{FK2012} blocking stack implementations based on the CC-Synch, DSM-Synch and H-Synch combining objects. Moreover, the lock-free stack implementation of \cite{T86} and the blocking implementation based on the CLH locks \cite{C93,MLH94} are provided.
The Synch framework also provides concurrent queue and stacks implementations (i.e. OsciQueue and OsciStack implementations) that achieve very high performance using user-level threads \cite{FK2017}.

Furthermore, the Synch framework provides a few scalable lock implementations, i.e. the MCS queue-lock presented in \cite{MCS91} and the CLH queue-lock presented in \cite{C93,MLH94}. Finally, the Synch framework provides two example-implementations of concurrent hash-tables. More specifically, it provides a simple implementation based on CLH queue-locks \cite{C93,MLH94} and an implementation based on the DSM-Synch \cite{FK2012} combining technique.

The following table presents a summary of the concurrent data-structures offered by the Synch framework.

\begin{table}[!h]
	\centering
\begin{tabular}{ll}
	\hline
	Concurrent  Object & Provided Implementations \\\hline	\hline
	Combining Objects &  CC-Synch, DSM-Synch and H-Synch \cite{FK2012}\\
	& PSim \cite{FK2011,FK2014} \\
	& Osci \cite{FK2017} \\
	& Oyama \cite{Oyama99} \\
	Concurrent Queues & CC-Queue, DSM-Queue and H-Queue \cite{FK2012} \\
	& SimQueue \cite{FK2011,FK2014} \\
	& OsciQueue \cite{FK2017} \\
	& CLH-Queue \cite{C93,MLH94} \\
	& MS-Queue \cite{MS96} \\
	Concurrent Stacks & CC-Stack, DSM-Stack and H-Stack \cite{FK2012} \\
	& SimStack \cite{FK2011,FK2014} \\
	& OsciStack \cite{FK2017} \\
	& CLH-Stack \cite{C93,MLH94} \\
	& LF-Stack \cite{T86} \\
	Locks & CLH \cite{C93,MLH94} \\
	& MCS \cite{MCS91}  \\
	Hash Tables & CLH-Hash \cite{C93,MLH94} \\
	& A hash-table based on DSM-Synch \cite{FK2012} \\
	\hline
\end{tabular}
\end{table}

\section{Benchmarks and performance optimizations}

For almost every concurrent data-structure, Synch provides at least one benchmark for evaluating its performance. The provided benchmarks allow users to assess the performance of concurrent data-structures, as well as to perform some basic correctness tests on them. All the provided benchmarks offer a great variety of command-line options for controlling the duration of the benchmark, the amount of processing cores and/or threads to be used, the contention, the type of threads (i.e. user-level or POSIX), etc.

\section{Source code structure}

\begin{figure}
	\centering
	\includegraphics[width=0.55\linewidth]{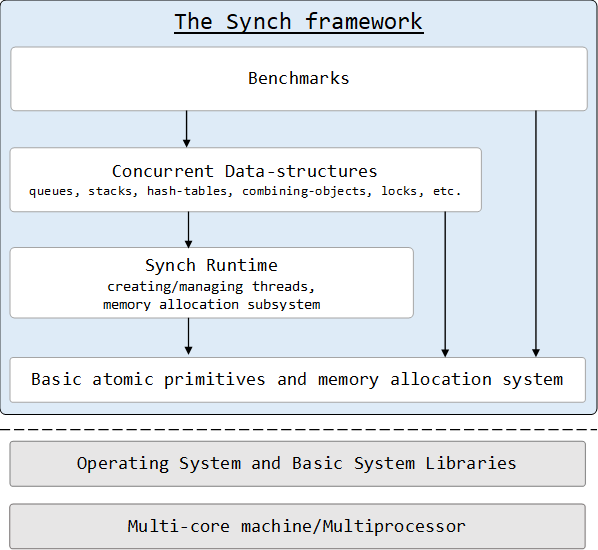}
	\caption{Code-structure of the Synch framework.}
	\label{fig:code_structure}
\end{figure}

The Synch framework (Figure \ref{fig:code_structure}) consists of 3 main parts, i.e. the Runtime/Primitives, the Concurrent library and the Benchmarks. The Runtime/Primitives part provides some basic functionality for creating and managing threads, functionality for basic atomic primitives (e.g. Compare\&Swap, Fetch\&Add, fences, simple synchronization barriers, etc.), mechanisms for memory allocation/management (e.g. memory pools, etc.), functionality for measuring time, reporting CPU counters, etc. Furthermore, the Runtime/Primitives provides a simple and lightweight library of user level-threads \cite{FK2017} that can be used in order to evaluate the provided data-structures and algorithms. The Concurrent library utilizes the building blocks of the Runtime/Primitives layer in order to provide all the concurrent data-structures (e.g. combining objects, queues, stacks, etc.). For almost every concurrent data-structure or synchronization mechanism, Synch provides at least one benchmark for evaluating its performance.

\section{Requirements}

\begin{itemize}
	\item A modern 64-bit multi-core machine. Currently, 32-bit architectures are not supported. The current version of this code is optimized for the x86\_64 machine architecture, but the code is also successfully tested in other machine architectures, such as ARM-V8 and RISC-V. For the case of x86\_64  architecture, the code has been evaluated in numerous Intel and AMD multicore machines. In the case of ARM-V8 architecture, the code has been successfully evaluated in a Trenz Zynq UltraScale+ board (4 A53 Cortex cores) and in a Raspberry Pi 3 board(4 Cortex A53 cores). For the RISC-V architecture, the code has been evaluated in a SiFive HiFive Unleashed (4 U54 RISC‑V cores) respectively. Some of the benchmarks may perform much better in architectures that natively support Fetch\&Add instructions (e.g., x86\_64, etc.).
	\item As a compiler, gcc of version 4.3 or greater is recommended, but the framework has been successfully built with icx and clang.
	\item Building requires the following development packages:
	\begin{itemize}
	\item \textit{libpapi} in the case that the user wants to measure performance using CPU performance counters.
	\item \textit{libnuma}
	\end{itemize}
\end{itemize}

\section{Acknowledgments}

This work was partially supported by the European Commission under the Horizon 2020 Framework Programme for Research and Innovation through the "European Processor Initiative: Specific Grant Agreement 1" (Grant Agreement Nr 826647).

Many thanks to Panagiota Fatourou for all the fruitful discussions and her significant contribution on the concurrent data-structures implementations presented in \cite{FK2011,FK2012,FK2014,FK2017}. 

Thanks also to Spiros Agathos for his feedback on the paper and committing some valuable patches to the repository. Many thanks also to Eftychia Datsika for her feedback on the paper.

\bibliographystyle{plain}
\bibliography{paper.bib}

\begin{thebibliography}{10}

\bibitem{AKD12}
Spiros~N. Agathos, Nikolaos~D. Kallimanis, and Vassilios~V. Dimakopoulos.
\newblock Speeding up openmp tasking.
\newblock In {\em Euro-Par 2012 Parallel Processing}, pages 650--661, Berlin,
  Heidelberg, 2012. Springer Berlin Heidelberg.

\bibitem{C93}
Travis~S. Craig.
\newblock Building fifo and priority queuing spin locks from atomic swap.
\newblock Technical report, University of Washington Computer Science
  Department, 1993.

\bibitem{FK2011}
Panagiota Fatourou and Nikolaos~D. Kallimanis.
\newblock A highly-efficient wait-free universal construction.
\newblock In {\em Proceedings of the Twenty-Third Annual ACM Symposium on
  Parallelism in Algorithms and Architectures}, SPAA '11, page 325–334, New
  York, NY, USA, 2011. Association for Computing Machinery.

\bibitem{FK2012}
Panagiota Fatourou and Nikolaos~D. Kallimanis.
\newblock Revisiting the combining synchronization technique.
\newblock In {\em Proceedings of the 17th ACM SIGPLAN Symposium on Principles
  and Practice of Parallel Programming}, PPoPP '12, page 257–266, New York,
  NY, USA, 2012. Association for Computing Machinery.

\bibitem{FK2014}
Panagiota Fatourou and Nikolaos~D Kallimanis.
\newblock Highly-efficient wait-free synchronization.
\newblock {\em Theory of Computing Systems}, 55(3):475--520, 2014.

\bibitem{FK2017}
Panagiota Fatourou and Nikolaos~D. Kallimanis.
\newblock {Lock Oscillation: Boosting the Performance of Concurrent Data
  Structures}.
\newblock In James Aspnes, Alysson Bessani, Pascal Felber, and Jo{\~a}o
  Leit{\~a}o, editors, {\em 21st International Conference on Principles of
  Distributed Systems (OPODIS 2017)}, volume~95 of {\em Leibniz International
  Proceedings in Informatics (LIPIcs)}, pages 8:1--8:17, Dagstuhl, Germany,
  2018. Schloss Dagstuhl--Leibniz-Zentrum fuer Informatik.

\bibitem{FKR2018}
Panagiota Fatourou, Nikolaos~D. Kallimanis, and Thomas Ropars.
\newblock An efficient wait-free resizable hash table.
\newblock In {\em Proceedings of the 30th on Symposium on Parallelism in
  Algorithms and Architectures}, SPAA '18, page 111–120, New York, NY, USA,
  2018. Association for Computing Machinery.

\bibitem{SynchFramework}
Nikolaos~D. Kallimanis.
\newblock {Synch: A framework for concurrent data-structures and benchmarks.
  https://github.com/nkallima/sim-universal-construction}.

\bibitem{MLH94}
P.~{Magnusson}, A.~{Landin}, and E.~{Hagersten}.
\newblock Queue locks on cache coherent multiprocessors.
\newblock In {\em Proceedings of 8th International Parallel Processing
  Symposium}, pages 165--171, 1994.

\bibitem{MCS91}
John~M. Mellor-Crummey and Michael~L. Scott.
\newblock Algorithms for scalable synchronization on shared-memory
  multiprocessors.
\newblock {\em ACM Trans. Comput. Syst.}, 9(1):21–65, February 1991.

\bibitem{MS96}
Maged~M. Michael and Michael~L. Scott.
\newblock Simple, fast, and practical non-blocking and blocking concurrent
  queue algorithms.
\newblock In {\em Proceedings of the Fifteenth Annual ACM Symposium on
  Principles of Distributed Computing}, PODC '96, page 267–275, New York, NY,
  USA, 1996. Association for Computing Machinery.

\bibitem{Oyama99}
Yoshihiro Oyama, Kenjiro Taura, and Akinori Yonezawa.
\newblock Executing parallel programs with synchronization bottlenecks
  efficiently.
\newblock In {\em Proceedings of the International Workshop on Parallel and
  Distributed Computing for Symbolic and Irregular Applications}, volume~16,
  page~95. Citeseer, 1999.

\bibitem{T86}
R~Kent Treiber.
\newblock Systems programming: Coping with parallelism.
\newblock Technical Report RJ-5118, International Business Machines
  Incorporated, Thomas J. Watson Research, 1986.

\end{thebibliography}

\end{document}